\documentclass[sigconf]{acmart}

\usepackage{url}

\newcommand{\ablConfBaseYelpClaude}{67.2}
\newcommand{\ablConfBaseYelpLlama}{75.4}
\newcommand{\ablConfPertYelpClaude}{53.0}
\newcommand{\ablConfPertYelpLlama}{85.6}

\newcommand{\ablEcePertYelpClaude}{0.380}

\newcommand{\ablGrndBaseYelpLlama}{87.1}
\newcommand{\ablGrndPertYelpClaude}{91.1}
\newcommand{\ablGrndPertYelpLlama}{98.5}

\newcommand{\ablOodBaseYelpLlama}{12.9}

\newcommand{\ablOodDropYelpLlama}{11.3}

\newcommand{\ablOodPertYelpLlama}{1.5}

\newcommand{\biasAlwaysInAmz}{+0.526}
\newcommand{\biasAlwaysInML}{+0.070}
\newcommand{\biasAlwaysInYelp}{+0.350}
\newcommand{\biasAlwaysIn}{+0.358}
\newcommand{\biasExactAmz}{-0.423}
\newcommand{\biasExactML}{-0.349}
\newcommand{\biasExactYelp}{-0.188}
\newcommand{\biasExact}{-0.313}
\newcommand{\biasLLMAmz}{-0.013}
\newcommand{\biasLLMML}{-0.023}
\newcommand{\biasLLMYelp}{-0.075}

\newcommand{\biasLLM}{-0.040}
\newcommand{\biasSpanAmz}{0.95}
\newcommand{\biasSurfaceAmz}{-0.372}
\newcommand{\biasSurfaceML}{-0.047}
\newcommand{\biasSurfaceYelp}{-0.163}
\newcommand{\biasSurface}{-0.219}
\newcommand{\biasTokenSetAmz}{+0.282}
\newcommand{\biasTokenSetML}{+0.070}
\newcommand{\biasTokenSetYelp}{+0.050}
\newcommand{\biasTokenSet}{+0.144}
\newcommand{\biasTokenSortAmz}{-0.372}
\newcommand{\biasTokenSortML}{-0.070}
\newcommand{\biasTokenSortYelp}{-0.175}
\newcommand{\biasTokenSort}{-0.229}
\newcommand{\blindAgreePct}{94.9}

\newcommand{\blindDisagree}{3}

\newcommand{\blindN}{59}
\newcommand{\blindPowerTen}{0.35}
\newcommand{\blindSlotsSame}{4}
\newcommand{\brierAmzClaudeHead}{0.311}
\newcommand{\brierAmzClaudeTail}{0.369}

\newcommand{\brierAmzGptOssHead}{0.414}
\newcommand{\brierAmzGptOssTail}{0.520}

\newcommand{\brierAmzLlamaHead}{0.361}
\newcommand{\brierAmzLlamaTail}{0.469}

\newcommand{\brierAmzMistralHead}{0.371}
\newcommand{\brierAmzMistralTail}{0.551}

\newcommand{\brierMLClaudeHead}{0.061}

\newcommand{\brierMLGptOssHead}{0.049}

\newcommand{\brierMLLlamaHead}{0.047}

\newcommand{\brierMLMistralHead}{0.035}

\newcommand{\brierYelpClaudeHead}{0.160}
\newcommand{\brierYelpClaudeTail}{0.170}

\newcommand{\brierYelpGptOssHead}{0.268}
\newcommand{\brierYelpGptOssTail}{0.294}

\newcommand{\brierYelpLlamaHead}{0.136}
\newcommand{\brierYelpLlamaTail}{0.154}

\newcommand{\brierYelpMistralHead}{0.111}
\newcommand{\brierYelpMistralTail}{0.104}

\newcommand{\cfAmzClaudeAfifteen}{50.7}
\newcommand{\cfAmzClaudeAfive}{50.5}
\newcommand{\cfAmzClaudeAten}{50.8}
\newcommand{\cfAmzClaudeAtwenty}{50.6}
\newcommand{\cfAmzGptOssAfifteen}{59.7}
\newcommand{\cfAmzGptOssAfive}{60.7}
\newcommand{\cfAmzGptOssAten}{60.5}
\newcommand{\cfAmzGptOssAtwenty}{59.4}
\newcommand{\cfAmzLlamaAfifteen}{53.5}
\newcommand{\cfAmzLlamaAfive}{53.7}
\newcommand{\cfAmzLlamaAten}{53.5}
\newcommand{\cfAmzLlamaAtwenty}{53.8}
\newcommand{\cfAmzMistralAfifteen}{49.6}
\newcommand{\cfAmzMistralAfive}{49.7}
\newcommand{\cfAmzMistralAten}{49.6}
\newcommand{\cfAmzMistralAtwenty}{50.0}
\newcommand{\cfEntries}{48}

\newcommand{\cfMaxRed}{1.65}
\newcommand{\cfMedRed}{0.02}
\newcommand{\cfWorseCells}{18}
\newcommand{\cfYelpClaudeAfifteen}{14.1}
\newcommand{\cfYelpClaudeAfive}{14.3}
\newcommand{\cfYelpClaudeAten}{14.2}
\newcommand{\cfYelpClaudeAtwenty}{14.0}
\newcommand{\cfYelpGptOssAfifteen}{37.9}
\newcommand{\cfYelpGptOssAfive}{38.2}
\newcommand{\cfYelpGptOssAten}{38.1}
\newcommand{\cfYelpGptOssAtwenty}{37.0}
\newcommand{\cfYelpLlamaAfifteen}{12.7}
\newcommand{\cfYelpLlamaAfive}{12.7}
\newcommand{\cfYelpLlamaAten}{12.7}
\newcommand{\cfYelpLlamaAtwenty}{12.9}
\newcommand{\cfYelpMistralAfifteen}{11.6}
\newcommand{\cfYelpMistralAfive}{11.9}
\newcommand{\cfYelpMistralAten}{11.8}
\newcommand{\cfYelpMistralAtwenty}{11.4}

\newcommand{\confBandHi}{86}
\newcommand{\confBandLo}{67}
\newcommand{\confBandWidth}{19}
\newcommand{\confMLClaude}{77.8}

\newcommand{\confMLMistral}{86.2}
\newcommand{\confModelClaude}{72.9}

\newcommand{\confModelMistral}{84.8}

\newcommand{\costCalls}{1,884}
\newcommand{\costUSD}{12}
\newcommand{\costVerdicts}{18,833}

\newcommand{\eceAmzClaudeHead}{0.238}
\newcommand{\eceAmzClaudeTail}{0.349}

\newcommand{\eceAmzGptOssHead}{0.418}
\newcommand{\eceAmzGptOssTail}{0.571}

\newcommand{\eceAmzLlamaHead}{0.331}
\newcommand{\eceAmzLlamaTail}{0.496}

\newcommand{\eceAmzMistralHead}{0.342}
\newcommand{\eceAmzMistralTail}{0.588}

\newcommand{\eceEqCells}{10}
\newcommand{\eceMLClaudeHead}{0.210}

\newcommand{\eceMLGptOssHead}{0.139}

\newcommand{\eceMLLlamaHead}{0.196}

\newcommand{\eceMLMistralHead}{0.118}

\newcommand{\eceYelpClaudeHead}{0.185}
\newcommand{\eceYelpClaudeTail}{0.193}

\newcommand{\eceYelpGptOssHead}{0.182}
\newcommand{\eceYelpGptOssTail}{0.224}

\newcommand{\eceYelpLlamaHead}{0.120}
\newcommand{\eceYelpLlamaTail}{0.112}

\newcommand{\eceYelpMistralHead}{0.066}
\newcommand{\eceYelpMistralTail}{0.098}

\newcommand{\errCatAmz}{0.218}
\newcommand{\errCatML}{0.023}
\newcommand{\errCatP}{0.0008}
\newcommand{\errCatYelp}{0.075}
\newcommand{\errFN}{16}
\newcommand{\errFP}{8}
\newcommand{\errGross}{24}
\newcommand{\fcHumanAmz}{48.9}

\newcommand{\fcTsetAmz}{93.8}

\newcommand{\foneAlwaysIn}{0.782}

\newcommand{\foneExact}{0.656}

\newcommand{\foneLLMYelp}{0.939}
\newcommand{\foneLLMexLeak}{0.902}

\newcommand{\foneLLM}{0.904}
\newcommand{\foneSpan}{0.25}

\newcommand{\foneSurface}{0.785}

\newcommand{\foneTokenSetYelp}{0.944}
\newcommand{\foneTokenSet}{0.857}

\newcommand{\foneTokenSort}{0.755}
\newcommand{\gapAmzClaude}{-24.4}
\newcommand{\gapAmzGptOss}{-42.7}

\newcommand{\grndAmzClaudeHead}{50.0}
\newcommand{\grndAmzClaudeTail}{40.0}

\newcommand{\grndAmzGptOss}{39.0}

\newcommand{\grndAmzMistralHead}{52.1}
\newcommand{\grndAmzMistralTail}{28.2}
\newcommand{\grndAmzMistral}{50.7}

\newcommand{\grndMLGptOss}{97.3}

\newcommand{\grndMLLlama}{99.4}

\newcommand{\grndRatioApprox}{2.6}
\newcommand{\grndSwing}{60}

\newcommand{\humanOutAmz}{41}

\newcommand{\kappaAlwaysIn}{0.000}

\newcommand{\kappaExact}{0.392}
\newcommand{\kappaLLMHi}{0.836}
\newcommand{\kappaLLMLo}{0.647}

\newcommand{\kappaLLM}{0.747}

\newcommand{\kappaSurface}{0.561}
\newcommand{\kappaTokenSetHi}{0.632}
\newcommand{\kappaTokenSetLo}{0.386}
\newcommand{\kappaTokenSet}{0.513}

\newcommand{\kappaTokenSort}{0.507}
\newcommand{\mcnemarP}{0.02}

\newcommand{\mmxAlwaysIn}{0.526}

\newcommand{\mmxLLM}{0.075}

\newcommand{\mmxTokenSet}{0.282}

\newcommand{\nAmzGptOss}{2947}

\newcommand{\nAmzVal}{78}
\newcommand{\nDrawn}{205}

\newcommand{\nLabels}{201}

\newcommand{\nMLMistral}{3000}
\newcommand{\nMLOut}{3}
\newcommand{\nOver}{5}
\newcommand{\nUnder}{7}

\newcommand{\nUsersAmzGptOss}{295}

\newcommand{\oodAmzHi}{60.7}
\newcommand{\oodAmzLo}{41.5}
\newcommand{\oodAmz}{51.1}
\newcommand{\oodCAmzClaudeHi}{52.9}
\newcommand{\oodCAmzClaudeLo}{48.1}
\newcommand{\oodCAmzClaude}{50.6}
\newcommand{\oodCAmzGptOssHi}{63.3}
\newcommand{\oodCAmzGptOssLo}{58.7}
\newcommand{\oodCAmzGptOss}{61.0}
\newcommand{\oodCAmzLlamaHi}{56.0}
\newcommand{\oodCAmzLlamaLo}{51.4}
\newcommand{\oodCAmzLlama}{53.6}
\newcommand{\oodCAmzMistralHi}{51.8}
\newcommand{\oodCAmzMistralLo}{46.7}
\newcommand{\oodCAmzMistral}{49.3}
\newcommand{\oodCMLClaudeHi}{1.7}
\newcommand{\oodCMLClaudeLo}{0.8}
\newcommand{\oodCMLClaude}{1.2}
\newcommand{\oodCMLGptOssHi}{3.6}
\newcommand{\oodCMLGptOssLo}{1.9}
\newcommand{\oodCMLGptOss}{2.7}
\newcommand{\oodCMLLlamaHi}{0.9}
\newcommand{\oodCMLLlamaLo}{0.4}
\newcommand{\oodCMLLlama}{0.6}
\newcommand{\oodCMLMistralHi}{2.6}
\newcommand{\oodCMLMistralLo}{1.4}
\newcommand{\oodCMLMistral}{2.0}
\newcommand{\oodCYelpClaudeHi}{16.2}
\newcommand{\oodCYelpClaudeLo}{12.7}
\newcommand{\oodCYelpClaude}{14.4}
\newcommand{\oodCYelpGptOssHi}{42.0}
\newcommand{\oodCYelpGptOssLo}{35.4}
\newcommand{\oodCYelpGptOss}{38.7}
\newcommand{\oodCYelpLlamaHi}{14.9}
\newcommand{\oodCYelpLlamaLo}{11.0}
\newcommand{\oodCYelpLlama}{12.9}
\newcommand{\oodCYelpMistralHi}{13.5}
\newcommand{\oodCYelpMistralLo}{9.7}
\newcommand{\oodCYelpMistral}{11.6}
\newcommand{\oodMLHi}{7.2}
\newcommand{\oodMLLo}{0.6}
\newcommand{\oodML}{1.2}
\newcommand{\oodYelpHi}{17.0}
\newcommand{\oodYelpLo}{8.4}
\newcommand{\oodYelp}{11.0}

\newcommand{\precAlwaysIn}{0.642}

\newcommand{\precExact}{0.970}
\newcommand{\precLLMAmz}{0.778}

\newcommand{\precLLM}{0.934}

\newcommand{\precSurface}{0.988}
\newcommand{\precTokenSetAmz}{0.542}

\newcommand{\precTokenSet}{0.778}

\newcommand{\precTokenSort}{0.964}

\newcommand{\recAlwaysIn}{1.000}

\newcommand{\recExactPct}{49.6}

\newcommand{\recExact}{0.496}

\newcommand{\recLLM}{0.876}

\newcommand{\recSurface}{0.651}

\newcommand{\recTokenSet}{0.953}

\newcommand{\recTokenSort}{0.620}

\newcommand{\rgBadML}{3}

\newcommand{\rgHiAmz}{65.4}

\newcommand{\rgLoAmz}{44.4}

\newcommand{\rgNOutML}{3}

\newcommand{\rwDLLM}{-2.5}
\newcommand{\rwDSurface}{-17.2}
\newcommand{\rwDTokenSet}{+10.1}
\newcommand{\rwDTokenSort}{-20.1}

\newcommand{\rwHuman}{78.9}

\newcommand{\spBiasClaudeAmz}{-0.071}
\newcommand{\spBiasClaude}{-0.041}
\newcommand{\spBiasOtherAmz}{+0.000}
\newcommand{\spBiasOther}{-0.039}

\newcommand{\spNClaudeAmz}{14}

\newcommand{\spPbias}{1.00}

\newcommand{\tsetFpAmz}{27}
\newcommand{\varBetweenCatalog}{6.0}
\newcommand{\varBetweenModel}{19.2}
\newcommand{\verifyAffected}{190}
\newcommand{\verifyBatches}{139}
\newcommand{\verifyExcessHi}{+5.0}
\newcommand{\verifyExcessLo}{-1.0}
\newcommand{\verifyExcess}{+2.0}
\newcommand{\verifyN}{1,390}
\newcommand{\verifyPartialBatches}{7}
\newcommand{\verifyP}{0.19}
\newcommand{\verifySampleIn}{500}
\newcommand{\verifySampleOut}{500}
\newcommand{\verifyUnansweredPct}{1.0}

\AtBeginDocument{%
  }

\setcopyright{none}
\renewcommand\footnotetextcopyrightpermission[1]{}
\copyrightyear{2026}
\acmYear{2026}
\acmConference[CIKM '26]{The 35th ACM International Conference on Information
  and Knowledge Management}{November 7--11, 2026}{Rome, Italy}

\begin{document}

\title{Do LLM Recommenders Know When They're Hallucinating?
  Auditing Confidence Calibration in Catalog Faithfulness}

\author{Srijith Ravikumar}
\authornote{This work was conducted independently and does not relate to the author's position or responsibilities at Amazon.}
\orcid{0009-0004-5928-9942}
\affiliation{%
  \institution{Amazon.com LLC}
  \city{Seattle}
  \state{Washington}
  \country{USA}
}
\email{srijith@amazon.com}

\begin{abstract}
LLM recommenders for top-$K$ item suggestion regularly emit titles outside the target catalog. Prior audits report a binary out-of-domain rate; none ask whether the model knew. We jointly audit hallucination rate (OOD@10) and verbalized-confidence calibration (ECE, Brier, reliability) for four zero-shot LLM recommenders from four independent vendors (Mistral Large, Llama-3.3-70B, GPT-OSS-120B, Claude Sonnet 4.6), not grounded or fine-tuned systems, across three catalogs (MovieLens-25M, Amazon Reviews 2023 Toys, Yelp Open Dataset), stratified by item popularity. Measuring catalog membership is itself the hard part: on identical outputs the reported rate moves by an order of magnitude with the string matcher used, and F1 cannot separate the candidates. We validate the instrument against \nLabels{} human judgments and select on net bias, where the adopted one is off by \biasLLM{} against \biasTokenSet{} for the common fuzzy rule. Hallucination is then strongly catalog-dependent (\oodCMLLlama--\oodCMLGptOss\% on MovieLens, \oodCYelpMistral--\oodCYelpGptOss\% on Yelp, \oodCAmzMistral--\oodCAmzGptOss\% on Amazon Toys). Each model holds a near-constant confidence level barely responsive to the catalog, while the catalog-hit rate swings \grndSwing{} points, so the sign of the error is set by where a model's constant lands against a catalog's accuracy: \nUnder{} of the twelve cells are under-confident and \nOver{} over-confident, all four under-confident on MovieLens, all four over-confident on Amazon Toys. We read this as an \emph{elicitation mismatch}: ``Just Ask'' elicits a generic quality rating, not a catalog-membership probability. A conformal abstention threshold over verbalized confidence changes hallucination by at most \cfMaxRed\,pp across four $\alpha$ levels, because the channel cannot separate correct items from hallucinations. We recommend that audits report calibration alongside OOD, validate the matcher producing the OOD number, and use catalog-anchored elicitation.
\end{abstract}

\ccsdesc[500]{Information systems~Recommender systems}
\ccsdesc[300]{Computing methodologies~Natural language generation}
\ccsdesc[300]{Information systems~Evaluation of retrieval results}

\keywords{Large Language Models, Recommender Systems, Hallucination,
  Confidence Calibration, Catalog Faithfulness, Selective Prediction}

\maketitle

\section{Introduction}
\label{sec:intro}

LLM-based recommenders are now in production for top-$K$ item suggestion across e-commerce, media, and local services~\cite{liao2025reclm,deng2025lcft,bao2024bigrec,shen2025alipay}. They also routinely emit titles that do not exist in the target catalog. The dominant response in the literature is engineering-driven grounding: RecLM-cgen~\cite{liao2025reclm}, logit-space constrained fine-tuning~\cite{deng2025lcft}, bi-step grounding~\cite{bao2024bigrec}, and retrieval rewriting~\cite{shen2025alipay} all aim to drive an out-of-domain rate, OOD@$K$, toward zero. The residual failures these systems still produce are reported as a single binary quantity. A grounded recommender that is \emph{equally confident} on its surviving hallucinations as on its correct items is indistinguishable from a calibrated one to any policy that gates on confidence. Recent calls for holistic generative-recommender evaluation flag exactly this gap~\cite{deldjoo2025holistic}.

Whether the model knew it was hallucinating is, empirically, a question about the joint distribution of the catalog-membership indicator and the model's verbalized confidence. We use ``know'' operationally throughout, referring to the model's verbalized confidence channel, without committing to a metaphysics of LLM knowledge. Two adjacent literatures have probed pieces of this joint distribution but not the joint itself. LLM-calibration work measures expected calibration error, Brier score, and reliability against open-ended QA oracles~\cite{tian2023justask,xiong2024canllms,kadavath2022mostlyknow,geng2024survey}. Recent theory connects miscalibration to hallucination as a population-level phenomenon~\cite{kalai2024calibrated,kalai2025whyhallucinate}. The closest neighbor in recommendation, Kweon et al.~\cite{kweon2025uncertainty}, decomposes uncertainty against \emph{ranking correctness}. To our knowledge, no prior work reports probabilistic calibration metrics jointly with catalog-faithfulness rate for LLM recommenders.

A methodological point precedes the empirical one. The membership indicator is produced by a string matcher, and that matcher is an instrument whose error is rarely reported. We find it cannot be taken for granted: on fixed model outputs, plausible off-the-shelf matchers disagree on the Amazon OOD rate by more than a factor of ten, because a permissive token-set rule counts a short or truncated catalog entry as a full match for any longer title containing its words, while a length-aware rule misses genuine alternate titles. We validate the matcher against \nLabels{} human judgments before using it, and find that the agreement statistic normally used to choose one cannot tell those matchers apart (\S\ref{sec:matcher}).

This paper makes three contributions. \textbf{First, a joint-audit protocol} for catalog-faithful LLM recommenders combining OOD@$K$, popularity-stratified calibration (ECE, Brier, reliability), and a split-conformal abstention threshold~\cite{yadkori2024conformal,kamath2020selective,wen2025knowyourlimits}, operationalized on three catalogs of contrasting size and canonicity and four instruction-tuned LLMs from four independent vendors, with the membership instrument itself validated rather than assumed. \textbf{Second, a diagnostic finding} that the binding constraint on inference-time abstention is the calibration quality of the verbalized-confidence channel: confidence sits in a \confBandWidth-point band while the catalog-hit rate varies by a factor of \grndRatioApprox{}, so the sign of its error is fixed by where a model-specific confidence level lands against a catalog-specific accuracy. \textbf{Third, an empirical ceiling} on verbalized-confidence-driven abstention: across our $\alpha$ sweep, hallucination changes by at most \cfMaxRed\,pp. All three findings are for zero-shot frontier LLMs used directly as recommenders, not for grounded, fine-tuned, or retrieval-constrained stacks.

\section{Related Work}
\label{sec:related}
{\noindent\textbf{LLM-recommender hallucination.}
A growing line of work targets out-of-catalog generation in LLM recommenders. RecLM~\cite{liao2025reclm} unifies three grounding strategies; LCFT~\cite{deng2025lcft} adds a KL-divergence term over positive and negative instruction pairs to separate them in logit space; bi-step grounding~\cite{bao2024bigrec} retrieves candidates before generating; the Alipay search system~\cite{shen2025alipay} rewrites generative retrieval queries to suppress hallucination at scale; ETEGRec~\cite{liu2025etegrec} learns end-to-end item tokens; and TIGER~\cite{rajput2023tiger} encodes items via semantic IDs. Across these systems, the dominant evaluation protocol reports a binary in-catalog rate. It does not report a faithfulness distribution conditioned on confidence, nor the error of the matcher producing that rate. Holistic-evaluation proposals~\cite{deldjoo2025holistic} argue the protocol is too coarse for generative recommenders, but do not themselves report calibration. Closest to our methodological point, Huang et al.~\cite{huang2026entitymatching} build a human-annotated benchmark for entity matching in recommender datasets and evaluate rule-, lexical-, embedding- and LLM-based matchers against it. Their task is cross-dataset linking, where a correct match almost always exists; ours is the hallucination decision, where the negative case is the quantity of interest, and we measure how far a published hallucination rate moves under matcher choice.

\vspace{0.2em}
\noindent\textbf{LLM calibration, selective prediction, and abstention.}
The calibration literature elicits confidence by token logprobs~\cite{kadavath2022mostlyknow}, verbalized self-assessment~\cite{tian2023justask}, or structured prompting~\cite{xiong2024canllms,geng2024survey}. Semantic entropy~\cite{farquhar2024semantic} provides a generation-internal hallucination signal. Selective prediction and abstention give bounded-error guarantees under shift~\cite{kamath2020selective,yadkori2024conformal,wen2025knowyourlimits,zhang2024rtuning,kirichenko2025abstentionbench}. None of this work has been applied to recommendation catalog faithfulness.

\vspace{0.2em}
\noindent\textbf{Closest neighbor.}
Closest in spirit, Kweon et al.~\cite{kweon2025uncertainty} decompose epistemic and aleatoric uncertainty for LLM recommenders. Their oracle is \emph{ranking correctness}; ours is \emph{catalog membership}, and a model can correlate uncertainty with ranking quality while shipping titles that do not exist. We also measure hallucination jointly with confidence and add a conformal abstention threshold~\cite{yadkori2024conformal}.\par}

\section{Methodology}
\label{sec:method}

\begin{table*}[t]
\centering
\small
\caption{Six candidate membership instruments scored against \nLabels{} human judgments, each on the eight retrieved candidates the annotator saw. \emph{Net bias} is the instrument's in-catalog rate minus the annotator's on the same items. F1 spans \foneSpan{} across these six while their net bias on Amazon spans \biasSpanAmz{}, so F1 cannot carry the selection and bias can. The token-set rule is the matcher most commonly used in this literature: on Amazon it counts \tsetFpAmz{} of the \humanOutAmz{} items the annotator called non-members as members, for precision \precTokenSetAmz{} against \precLLMAmz{} for the adopted judge.}
\label{tab:matcher}
\begin{tabular}{@{}lcccc cccc@{}}
\toprule
 & & & & & \multicolumn{4}{c}{\textbf{Net bias}} \\
\cmidrule(l){6-9}
\textbf{Instrument} & \textbf{Prec.} & \textbf{Rec.} & \textbf{F1} & $\kappa$ &
\textbf{MovieLens} & \textbf{Yelp} & \textbf{Amazon} & \textbf{All} \\
\midrule
Always in catalog (trivial) & \precAlwaysIn & \recAlwaysIn & \foneAlwaysIn & \kappaAlwaysIn & \biasAlwaysInML & \biasAlwaysInYelp & \biasAlwaysInAmz & \biasAlwaysIn \\
Exact match after folding   & \precExact & \recExact & \foneExact & \kappaExact & \biasExactML & \biasExactYelp & \biasExactAmz & \biasExact \\
Token-set $\geq$ 90 (published) & \precTokenSet  & \recTokenSet  & \foneTokenSet & \kappaTokenSet & \biasTokenSetML & \biasTokenSetYelp & \biasTokenSetAmz & \biasTokenSet \\
Token-sort $\geq$ 90        & \precTokenSort & \recTokenSort & \foneTokenSort & \kappaTokenSort & \biasTokenSortML & \biasTokenSortYelp & \biasTokenSortAmz & \biasTokenSort \\
Surface-form rules          & \precSurface   & \recSurface   & \foneSurface & \kappaSurface & \biasSurfaceML & \biasSurfaceYelp & \biasSurfaceAmz & \biasSurface \\
LLM judge (adopted)         & \precLLM       & \recLLM       & \foneLLM & \kappaLLM & \biasLLMML & \biasLLMYelp & \biasLLMAmz & \biasLLM \\
\bottomrule
\end{tabular}
\end{table*}

\subsection{Datasets and catalog construction}
We audit three catalogs spanning canonical entertainment, long-tail e-commerce, and regional services. MovieLens-25M~\cite{movielens25m} contributes 62{,}423 movies with rich title and year metadata that recurs heavily in LLM pre-training data. Amazon Reviews 2023 Toys \& Games~\cite{hou2024amazonreviews} contributes 890{,}874 product titles, the large majority of which are obscure long-tail items. Yelp Open Dataset~\cite{yelp_open_dataset} contributes 150{,}346 businesses keyed by name and city, with strong regional concentration. For each catalog we construct an audit set of 300 users with at least 30 prior interactions, plus a disjoint held-out set of 100 users for conformal calibration. Per-cell counts are \nUsersAmzGptOss--300 users and \nAmzGptOss--\nMLMistral{} recommendations after dropping errored or unparseable responses; only GPT-OSS on Amazon and Yelp falls below 300 users. Users are grouped by the popularity quartile of their held-out target item, and we report head (Q1--Q2) and long-tail (Q3--Q4) statistics in \S\ref{sec:findings}.

\subsection{LLMs and prompting protocol}
We evaluate four instruction-tuned LLMs from four independent vendors: Mistral Large (\texttt{mistral-\allowbreak large-\allowbreak latest}, via La Plateforme), Llama-3.3-70B-Instruct and GPT-OSS-120B (\texttt{meta-\allowbreak llama/\allowbreak Llama-\allowbreak 3.3-\allowbreak 70B-\allowbreak Instruct-\allowbreak Turbo} and \texttt{openai/\allowbreak gpt-\allowbreak oss-\allowbreak 120b}, both via Together AI), and Claude Sonnet 4.6 (\texttt{claude-\allowbreak sonnet-\allowbreak 4-6}, Anthropic), all accessed 2026-05-08 to 2026-05-09. Each receives a zero-shot prompt containing the user's 30 most recent interactions and produces a top-10 list. Decoding is held fixed across models (temperature 0, $K=10$, a single fixed system prompt). We refer to a (catalog, LLM) pair as a \emph{cell}; there are twelve.

\subsection{Measuring catalog membership, and validating the measurement}
\label{sec:matcher}
A generated title counts as in-catalog if it refers to the same real-world item as some catalog entry. Exact matching is too strict, because models emit alternate titles, moved articles, and accent variants for items that exist: after Unicode folding it recovers only \recExactPct\% of the items our annotator confirmed. Fuzzy matching is the usual remedy, and its choice turns out to dominate the reported rate. A token-set rule at threshold 90, the most common off-the-shelf choice, returns a perfect score whenever one token set contains the other, so a truncated catalog entry matches any longer title containing its words. A length-aware token-sort rule at the same threshold removes that failure but introduces the opposite one, missing alternate titles such as \emph{Se7en} against the MovieLens entry \emph{Seven (a.k.a. Se7en) (1995)}.

We treat the matcher as an instrument to be measured. The usual practice is to select an LLM judge by its agreement with human labels~\cite{thomas2024llmjudge} and validate it by whether it preserves system orderings~\cite{faggioli2023perspectives}; neither check constrains the rate the instrument reports, which is the quantity a hallucination audit publishes. From a frame of 1{,}200 recommendations judged for membership we drew \nDrawn{} items stratified on that judgment, and hand-labeled \nLabels{} of them; four were left undecided and dropped, two of them from the small MovieLens out-of-catalog stratum, which leaves \nMLOut{} rows there. Each item was shown with eight candidate catalog entries, the same number the deployed matcher sees, and the annotator marked which candidate, if any, denotes the same item. Independent retrieval matters: an earlier round drew candidates from the matcher under test, which made that matcher's own false negatives unobservable and returned a recall estimate of exactly 1.0 on the catalog where it errs most. Those labels were discarded. Candidates now come from an index that unions an IDF-weighted rare-token block with a fuzzy-ranked block. Only the first block is independent of the rules under test: the second computes the same similarity the surface rules do, so the retrieval is not a fully independent second opinion, and the gold standard inherits whatever that index misses.

Table~\ref{tab:matcher} scores six instruments, including two trivial baselines. F1 is the conventional criterion and it barely separates them: it runs \foneExact--\foneLLM{}, and a predictor that simply calls every title in-catalog scores \foneAlwaysIn{}, above one of the four real matchers and level with a second. That classification accuracy and class-prevalence accuracy come apart is the founding observation of quantification learning~\cite{forman2008quantifying,jerzak2023quantification}, shown for LLM judges specifically by Dorner et al.~\cite{dorner2025limits}; we import the criterion rather than propose it. What a prevalence estimate is sensitive to is net bias, the gap between the in-catalog rate an instrument reports and the rate the annotator gives on the same items. On Amazon that gap runs from \biasExactAmz{} to \biasAlwaysInAmz{}, a spread of \biasSpanAmz{} against \foneSpan{} for F1. We select on bias. The adopted instrument is an LLM judge (\texttt{claude-sonnet-4-6}, temperature 0) given the same eight candidates and asked which, if any, denotes the same item; its bias is \biasLLM{} overall and \biasLLMAmz{} on Amazon, where the smallest error among the alternatives is \biasTokenSetAmz{}. Reweighted to the sampling frame, the annotator puts \rwHuman\% of recommendations in catalog; the judge lands \rwDLLM{} points away, the token-set rule \rwDTokenSet{}, the surface rules \rwDSurface{}, token-sort \rwDTokenSort{}. Cohen's $\kappa$ against the human labels is \kappaLLM{} [\kappaLLMLo, \kappaLLMHi] for the judge and \kappaTokenSet{} [\kappaTokenSetLo, \kappaTokenSetHi] for the token-set rule.

Table~\ref{tab:matcher} scores every instrument on the eight candidates, the one information set they all share. In practice the token-set rule is run over the entire catalog, which makes it more permissive still: on these same Amazon items it then puts \fcTsetAmz\% in catalog, where the annotator puts \fcHumanAmz\%. Its bias column is a lower bound on the bias of the configuration the literature actually uses.

We also show what correcting for that error does. Rogan-Gladen rescaling~\cite{rogan1978estimating} by per-catalog sensitivity and specificity moves the Amazon cells to \rgLoAmz--\rgHiAmz\% and leaves the catalog ordering intact, but is unusable where specificity rests on few labels: with \rgNOutML{} out-of-catalog labels on MovieLens, \rgBadML{} of its four corrected cells fall below zero. The raw rates stay primary. Stated properly the criterion is the worst per-catalog $|$net bias$|$, which is \mmxLLM{} for the judge against \mmxTokenSet{} to \mmxAlwaysIn{} for the five alternatives. Net bias hides a sum: the judge makes \errFP{} false positives and \errFN{} false negatives, so \errGross{} errors net to \biasLLM{} of the rate. The cancellation is what the criterion is for, a prevalence estimate being sensitive only to the difference, but it bounds the reported rate and not the per-item reliability. The judge does not dominate. It loses Yelp to the token-set rule on F1 (\foneLLMYelp{} against \foneTokenSetYelp) and on net bias (\biasLLMYelp{} against \biasTokenSetYelp), which is the criterion we selected on, so the concession is the one that counts. Its accuracy advantage over that rule is significant (McNemar $p = \mcnemarP$), but accuracy is not what the selection rests on. Its own error rate follows the same catalog gradient the paper reports: \errCatML{} on MovieLens, \errCatYelp{} on Yelp, \errCatAmz{} on Amazon (permutation $p = \errCatP$). An instrument that errs on a quarter of Amazon items cannot manufacture a gap of \fcTsetAmz\% against \fcHumanAmz\%, but it does mean the Amazon figure should be read as the interval in \S\ref{sec:discussion} rather than as a point.

An item omitted from a batched reply was recorded as a non-match, measured at \verifyUnansweredPct\% of items and quantified below. Three threats to the validation itself are worth stating. The judge shares a vendor with one of the four generators. It never sees which model produced a title, and its net bias on Claude-generated items is \spBiasClaude{} against \spBiasOther{} on the other three (permutation $p = \spPbias$). Pooling is the wrong test where judge error is catalog-dependent, and restricted to Amazon the figures are \spBiasClaudeAmz{} against \spBiasOtherAmz{} on \spNClaudeAmz{} items, stricter on its own vendor's output. The direction is against the objection either way, and neither reading is powered to settle it. The labeling form displayed the judge's verdict, which anchors toward it. We relabeled \blindN{} items with that column hidden and reproduce the original labels on \blindAgreePct\% of them, but that pass does not measure anchoring: it drew its candidates at a shallower depth, so only the first \blindSlotsSame{} of the eight shown entries are common to both forms, and all \blindDisagree{} disagreements are picks in the slots that changed. Hint removal and candidate substitution are fully confounded there. At this sample size the design has power \blindPowerTen{} against a ten-point effect, so what we can say is that the two label sets agree closely, not that anchoring was absent. Finally, an LLM judge validating a pipeline that also uses an LLM invites a circularity objection~\cite{faggioli2023perspectives,zheng2023judging}: our defense is that the judge is anchored to human labels instead of standing in for them, that we report its residual error, and that it selects among retrieved catalog entries rather than scoring free model output.

\subsection{Confidence elicitation}
We use verbalized self-rated confidence on a 0--100 scale, elicited via the Tian et al.~\cite{tian2023justask} ``Just Ask'' template, applied uniformly across all LLMs. Verbalized confidence is the natural elicitation for production deployment: it works uniformly across closed-source APIs and open-weight models, and Tian et al.\ document that token logprobs on RLHF'd LLMs are systematically over-confident relative to verbalized self-rating. Cross-vendor replication across the LLMs in our audit serves as the cross-channel robustness control. We also run catalog-anchored prompt ablations on Yelp (\S\ref{sec:findings}).

\subsection{Calibration metrics and conformal threshold}
\begin{figure}[t]
\centering
\includegraphics[width=\columnwidth]{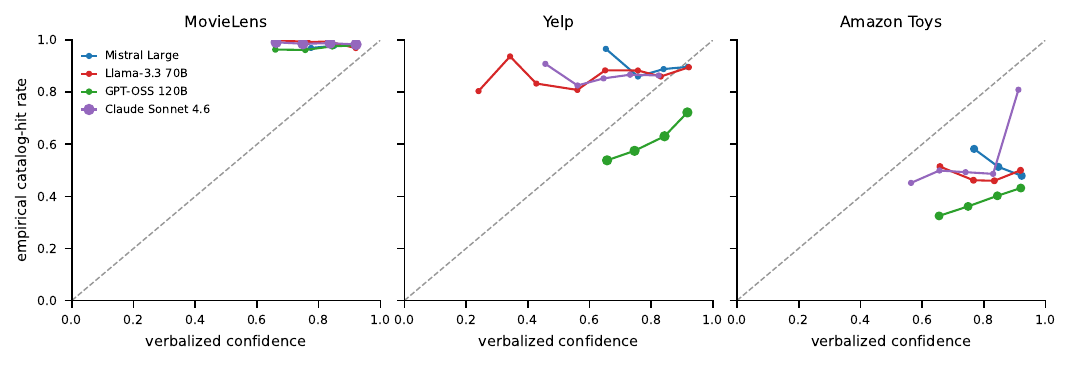}
\caption{Reliability per catalog, all four LLMs on the overall stratum. Gray dashed: perfect calibration. Curves sit above the diagonal on MovieLens (under-confidence) and below it on Amazon Toys (over-confidence), with Yelp mixed. The direction of miscalibration is a property of the catalog, not of the model.}
\Description{Three side-by-side reliability panels, one per catalog (MovieLens, Yelp, Amazon Toys). Each plots empirical catalog-hit rate against verbalized confidence for four LLMs: Mistral Large in blue, Llama-3.3 70B in red, GPT-OSS 120B in green, Claude Sonnet 4.6 in purple, against a gray dashed perfect-calibration diagonal. On MovieLens all plotted bins lie above the diagonal; on Amazon Toys all lie below it; Yelp contains bins on both sides.}
\label{fig:reliability}
\end{figure}

We report Expected Calibration Error (ECE) using 10-bin equal-width binning over normalized verbalized confidence, and Brier score alongside it as a binning-free check. Reliability diagrams (Figure~\ref{fig:reliability}) are drawn per cell. For the conformal threshold we follow split-conformal calibration~\cite{yadkori2024conformal}. Let $c_i \in [0,1]$ denote normalized verbalized confidence on calibration item $i$, and define the nonconformity score $s_i = 1 - c_i$. We take $\hat{q}_\alpha$ as the $\lceil (n+1)(1-\alpha) \rceil / n$ empirical quantile of $\{s_i\}$ over \emph{in-catalog} calibration items. That buys a retention guarantee on in-catalog items, not marginal coverage over the mixture, and it is a statement about the score quantile rather than about hallucination probability: transferring it needs $s_i$ to be a calibrated hallucination indicator, which \S\ref{sec:findings} shows fails. \S\ref{sec:mitigation} reports the empirical retained-set rates.

\section{Findings}
\label{sec:findings}

\begin{table}[t]
\centering
\small
\setlength{\tabcolsep}{4pt}
\caption{OOD@10 hallucination rate under the adopted instrument (mean over audit users; 95\% paired-user bootstrap CI in brackets).}
\label{tab:ood}
\begin{tabular}{@{}lccc@{}}
\toprule
\textbf{LLM} & \textbf{MovieLens} & \textbf{Amazon Toys} & \textbf{Yelp} \\
\midrule
Mistral Large     & \oodCMLMistral & \oodCAmzMistral & \oodCYelpMistral \\
                  & {\scriptsize [\oodCMLMistralLo,\oodCMLMistralHi]} & {\scriptsize [\oodCAmzMistralLo,\oodCAmzMistralHi]} & {\scriptsize [\oodCYelpMistralLo,\oodCYelpMistralHi]} \\
Llama-3.3 70B     & \oodCMLLlama & \oodCAmzLlama & \oodCYelpLlama \\
                  & {\scriptsize [\oodCMLLlamaLo,\oodCMLLlamaHi]} & {\scriptsize [\oodCAmzLlamaLo,\oodCAmzLlamaHi]} & {\scriptsize [\oodCYelpLlamaLo,\oodCYelpLlamaHi]} \\
GPT-OSS 120B      & \oodCMLGptOss & \oodCAmzGptOss & \oodCYelpGptOss \\
                  & {\scriptsize [\oodCMLGptOssLo,\oodCMLGptOssHi]} & {\scriptsize [\oodCAmzGptOssLo,\oodCAmzGptOssHi]} & {\scriptsize [\oodCYelpGptOssLo,\oodCYelpGptOssHi]} \\
Claude Sonnet 4.6 & \oodCMLClaude & \oodCAmzClaude & \oodCYelpClaude \\
                  & {\scriptsize [\oodCMLClaudeLo,\oodCMLClaudeHi]} & {\scriptsize [\oodCAmzClaudeLo,\oodCAmzClaudeHi]} & {\scriptsize [\oodCYelpClaudeLo,\oodCYelpClaudeHi]} \\
\bottomrule
\end{tabular}
\end{table}

\subsection{Hallucination is catalog-dependent}
OOD@10 in Table~\ref{tab:ood} traces a steep gradient by catalog \emph{canonicity}: \oodCMLLlama--\oodCMLGptOss\% on MovieLens, \oodCYelpMistral--\oodCYelpGptOss\% on Yelp, and \oodCAmzMistral--\oodCAmzGptOss\% on Amazon Toys. Our hypothesis is memorization at pre-training scale: MovieLens titles are canonical film entries seen many times in any web-scale corpus, Amazon Toys are long-tail product strings rarely surfaced outside retail pages, and Yelp couples business names to U.S. cities. The model reproduces what it has memorized and confabulates the rest.

The gradient does not track parameter count: GPT-OSS-120B, the largest model in our lineup, sits at the high end on both long-tail catalogs, while Mistral, Llama, and Claude cluster on Amazon. We read the ordering between models cautiously. The instrument's own error rate varies by catalog, and we have no evidence it is independent of the generating model, so the separation we claim is between catalogs, where it is an order of magnitude, and not between models within a catalog, where the spread is a few points. The operational consequence is direct. Representation-level mitigation (RecLM~\cite{liao2025reclm}, LCFT~\cite{deng2025lcft}) is most valuable on the long-tail end of the canonicity gradient; on canonical catalogs at frontier scale, OOD@10 is already low and the binding evaluation axis is calibration.

\subsection{The direction of miscalibration is set by the catalog}
Verbalized confidence occupies a narrow band across every condition we test: cell means run \confBandLo--\confBandHi, a spread of \confBandWidth{} points. The catalog-hit rate over the same twelve cells runs \grndAmzGptOss--\grndMLLlama\%. The two quantities belong to different sources. Mean confidence is mostly a property of the model: between-model variance is \varBetweenModel{} against \varBetweenCatalog{} between catalogs, and a model carries its level wherever it is pointed, Claude at \confModelClaude{} and Mistral at \confModelMistral{}. Accuracy is a property of the catalog and swings \grndSwing{} points. The sign of the calibration error follows from where a model's near-constant confidence lands against a catalog's accuracy.

The consequence is visible in Figure~\ref{fig:reliability}. On MovieLens, where the catalog-hit rate is \grndMLGptOss--\grndMLLlama\%, every plotted bin lies above the diagonal and all four LLMs are under-confident, verbalizing \confMLClaude--\confMLMistral{} on items they get right \grndMLGptOss--\grndMLLlama\% of the time. On Amazon Toys, where the hit rate is \grndAmzGptOss--\grndAmzMistral\%, every bin lies below the diagonal and all four are over-confident, with gaps of \gapAmzClaude{} to \gapAmzGptOss{} points. Yelp sits between, with three LLMs under-confident and GPT-OSS over-confident. Across the twelve cells, \nUnder{} are under-confident and \nOver{} over-confident. The split is stable under instrument choice: of the twelve cells, eleven receive the same verdict under all three corrected matchers we tested, the exception being Mistral on Yelp, whose small positive gap is within the flat band under one of them. Two of those three agree by construction on the catalog that carries the headline: token-sort and the surface rules return identical predictions on all \nAmzVal{} Amazon validation items.

ECE and Brier appear in Table~\ref{tab:ece}. One caveat: where confidence concentrates in a few bins of one sign, ECE reduces numerically to the absolute mean gap, as it does in \eceEqCells{} of the twelve cells. Brier does not collapse and is the more informative of the two.

Why does the verbalized channel move so little? ``Just Ask'' elicits a generic recommendation-quality rating instead of a catalog-membership probability. The LLM has no separate mechanism to introspect membership and instead rates taste in the hedged band trained in by RLHF, while existence accuracy varies enormously by catalog. Two catalog-anchored prompt-perturbation ablations support this reading by moving the channel in \emph{opposite} directions on the same task. On Llama\,$\times$\,Yelp the perturbed prompt raises mean confidence \ablConfBaseYelpLlama\,$\to$\,\ablConfPertYelpLlama{} and raises the hit rate \ablGrndBaseYelpLlama\,$\to$\,\ablGrndPertYelpLlama\%, i.e. OOD@10 \ablOodBaseYelpLlama\,$\to$\,\ablOodPertYelpLlama\%, a \ablOodDropYelpLlama\,pp reduction against the \cfMaxRed\,pp ceiling of \S\ref{sec:mitigation}; the anchored prompt also changes the generated list, so this is not a pure confidence intervention; on Claude\,$\times$\,Yelp the same prompt lowers mean confidence \ablConfBaseYelpClaude\,$\to$\,\ablConfPertYelpClaude{} while the hit rate rises to \ablGrndPertYelpClaude\%, widening the reliability gap and raising ECE to \ablEcePertYelpClaude. Prompt phrasing materially moves the verbalized channel while catalog-hit rate moves far less, and its magnitude and direction are template- and vendor-specific.

\begin{table}[t]
\centering
\small
\setlength{\tabcolsep}{3pt}
\caption{ECE (10-bin equal-width) and Brier score by LLM, dataset and popularity stratum of the user's target item. MovieLens tail strata fall below the 50-item threshold and are omitted. Lower is better.}
\label{tab:ece}
\begin{tabular}{@{}llcccc@{}}
\toprule
\textbf{LLM} & \textbf{Dataset} & \textbf{ECE-head} & \textbf{ECE-tail} & \textbf{Brier-h.} & \textbf{Brier-t.} \\
\midrule
Mistral & MovieLens & \eceMLMistralHead & --- & \brierMLMistralHead & --- \\
Mistral & Yelp      & \eceYelpMistralHead & \eceYelpMistralTail & \brierYelpMistralHead & \brierYelpMistralTail \\
Mistral & Amazon    & \eceAmzMistralHead & \eceAmzMistralTail & \brierAmzMistralHead & \brierAmzMistralTail \\
\addlinespace
Llama   & MovieLens & \eceMLLlamaHead & --- & \brierMLLlamaHead & --- \\
Llama   & Yelp      & \eceYelpLlamaHead & \eceYelpLlamaTail & \brierYelpLlamaHead & \brierYelpLlamaTail \\
Llama   & Amazon    & \eceAmzLlamaHead & \eceAmzLlamaTail & \brierAmzLlamaHead & \brierAmzLlamaTail \\
\addlinespace
GPT-OSS & MovieLens & \eceMLGptOssHead & --- & \brierMLGptOssHead & --- \\
GPT-OSS & Yelp      & \eceYelpGptOssHead & \eceYelpGptOssTail & \brierYelpGptOssHead & \brierYelpGptOssTail \\
GPT-OSS & Amazon    & \eceAmzGptOssHead & \eceAmzGptOssTail & \brierAmzGptOssHead & \brierAmzGptOssTail \\
\addlinespace
Claude  & MovieLens & \eceMLClaudeHead & --- & \brierMLClaudeHead & --- \\
Claude  & Yelp      & \eceYelpClaudeHead & \eceYelpClaudeTail & \brierYelpClaudeHead & \brierYelpClaudeTail \\
Claude  & Amazon    & \eceAmzClaudeHead & \eceAmzClaudeTail & \brierAmzClaudeHead & \brierAmzClaudeTail \\
\bottomrule
\end{tabular}
\end{table}

\subsection{Miscalibration worsens on the long tail}
Splitting users by the popularity quartile of their target item shows the effect concentrating where visibility is lowest. On Amazon the catalog-hit rate falls from \grndAmzMistralHead\% to \grndAmzMistralTail\% for Mistral and from \grndAmzClaudeHead\% to \grndAmzClaudeTail\% for Claude, while mean confidence is essentially unchanged, so tail ECE rises to \eceAmzMistralTail{} and \eceAmzClaudeTail{} respectively. Yelp shows the same direction at smaller magnitude. MovieLens tail strata contain too few users to support stratified inference and are omitted. Tail items are the least represented in pre-training, and the confidence channel does not register the difference.

\section{Calibration-Aware Filtering}
\label{sec:mitigation}

We sweep $\alpha \in \{0.05, 0.10, 0.15, 0.20\}$ in the conformal procedure of \S\ref{sec:method}; Table~\ref{tab:conformal} reports the two long-tail catalogs. Across all \cfEntries{} (cell, $\alpha$) entries the largest hallucination reduction is \cfMaxRed\,pp and the median is \cfMedRed\,pp; \cfWorseCells{} entries leave the retained set \emph{more} hallucinated than the unfiltered set. On Amazon, where roughly half the recommendations do not exist in the target catalog, discarding the least-confident tenth removes well under a point.

The asymmetry between what the procedure delivers and what an operator wants is the point. Retention behaves as the construction implies, but the retained set is no cleaner, because the score being thresholded carries almost no information about catalog membership. Any abstention policy built on verbalized confidence inherits that ceiling regardless of the procedure layered on top; lifting it requires either catalog-anchored elicitation or training-time objectives that target membership directly~\cite{zhang2024rtuning,farquhar2024semantic}.

\begin{table}[t]
\centering
\small
\setlength{\tabcolsep}{4pt}
\caption{Hallucination rate of the retained set under split-conformal abstention on verbalized confidence. MovieLens omitted (base rate below \oodCMLGptOss\%).}
\label{tab:conformal}
\begin{tabular}{@{}llccccc@{}}
\toprule
\textbf{Catalog} & \textbf{LLM} & \textbf{unfilt.} & $\alpha{=}.05$ & $\alpha{=}.10$ & $\alpha{=}.15$ & $\alpha{=}.20$ \\
\midrule
Yelp   & Mistral & \oodCYelpMistral & \cfYelpMistralAfive & \cfYelpMistralAten & \cfYelpMistralAfifteen & \cfYelpMistralAtwenty\\
Yelp   & Llama   & \oodCYelpLlama & \cfYelpLlamaAfive & \cfYelpLlamaAten & \cfYelpLlamaAfifteen & \cfYelpLlamaAtwenty\\
Yelp   & GPT-OSS & \oodCYelpGptOss & \cfYelpGptOssAfive & \cfYelpGptOssAten & \cfYelpGptOssAfifteen & \cfYelpGptOssAtwenty\\
Yelp   & Claude  & \oodCYelpClaude & \cfYelpClaudeAfive & \cfYelpClaudeAten & \cfYelpClaudeAfifteen & \cfYelpClaudeAtwenty\\
\addlinespace
Amazon & Mistral & \oodCAmzMistral & \cfAmzMistralAfive & \cfAmzMistralAten & \cfAmzMistralAfifteen & \cfAmzMistralAtwenty\\
Amazon & Llama   & \oodCAmzLlama & \cfAmzLlamaAfive & \cfAmzLlamaAten & \cfAmzLlamaAfifteen & \cfAmzLlamaAtwenty\\
Amazon & GPT-OSS & \oodCAmzGptOss & \cfAmzGptOssAfive & \cfAmzGptOssAten & \cfAmzGptOssAfifteen & \cfAmzGptOssAtwenty\\
Amazon & Claude  & \oodCAmzClaude & \cfAmzClaudeAfive & \cfAmzClaudeAten & \cfAmzClaudeAfifteen & \cfAmzClaudeAtwenty\\
\bottomrule
\end{tabular}
\end{table}

\section{Discussion and Limitations}
\label{sec:discussion}

All results are for zero-shot frontier LLMs used directly as recommenders. Grounded and fine-tuned stacks are out of scope, but the measurement point applies to them as directly, since they are audited with the same instruments.

The membership instrument carries residual error that we report rather than assume away: precision \precLLM{} and recall \recLLM{} against \nLabels{} labels from a single annotator, concentrated on numbered and sized product variants and spread unevenly across catalogs (\errCatML{} on MovieLens, \errCatAmz{} on Amazon). Per-catalog rates are best read as intervals: anchoring to the human labels gives \oodAmz\% [\oodAmzLo, \oodAmzHi] on Amazon, \oodYelp\% [\oodYelpLo, \oodYelpHi] on Yelp, and \oodML\% [\oodMLLo, \oodMLHi] on MovieLens. Table~\ref{tab:ood} is raw per-cell instrument output and need not fall inside these; GPT-OSS on Yelp does not.

\section*{GenAI Usage Disclosure}
The author used an AI coding assistant for experiment scaffolding, analysis code, and manuscript editing. All experimental design, the human validation labels, and all claims and conclusions are the author's own, and every reported number is generated directly from committed analysis artifacts.

An LLM also serves as the catalog-membership judge of \S\ref{sec:matcher}, which is a measurement instrument, not an editing aid, so we state its configuration in full. Model \texttt{claude-\allowbreak sonnet-\allowbreak 4-6} at temperature 0, prompted with a fixed system block and batches of ten items, each item carrying the generated title and eight candidate catalog entries; the judge returns a candidate index or zero. The pass produced \costVerdicts{} verdicts over \costCalls{} cached-prefix calls at an API cost of roughly \$\costUSD. Two of the \nLabels{} labeled items appear as examples in the system prompt; excluding them moves the judge's F1 from \foneLLM{} to \foneLLMexLeak. Hosted decoding at temperature 0 is not bit-reproducible, so a re-run will not reproduce every verdict exactly; the released verdict files are the ones all reported numbers derive from. One operational limit is worth stating because it biases in a known direction. The judge answers in batches of ten, and a reply occasionally omits an item: measured on a \verifyN-record re-judge, \verifyUnansweredPct\% of items went unanswered on the first pass, spread over \verifyPartialBatches{} of \verifyBatches{} batches. An unanswered item was recorded as a non-match, so of the \costVerdicts{} published verdicts on the order of \verifyAffected{} were never actually judged and are counted as hallucinations. Re-judging \verifySampleOut{} out-of-catalog records against \verifySampleIn{} in-catalog controls puts the excess flip rate at \verifyExcess\,pp (95\% CI \verifyExcessLo{} to \verifyExcessHi, $p = \verifyP$), so the effect on the rates in Table~\ref{tab:ood} is under a point and within their intervals. The human-anchored figures in \S\ref{sec:discussion} are unaffected, being anchored to the labels rather than to the instrument. The released pipeline now records whether each item was answered.

\vspace{0.3em}
\noindent\textbf{Data availability.}
The \costVerdicts{} judge verdicts, the \nLabels{} human labels with every instrument's verdict, the \blindN{} blind relabels, the annotation guideline, the verbatim judge prompt, and the scoring code that regenerates Table~\ref{tab:matcher} and the instrument statistics of \S\ref{sec:matcher} are available at \url{https://github.com/rsrijith/cikm26-catalog-faithfulness}. The MovieLens and Yelp licenses do not permit redistributing their catalogs, so entries from those two are named there by item identifier instead of by title, with a script that restores the titles from a reader's own licensed copy; we verified that the round trip returns every affected record unchanged. Amazon titles are included directly. The three source catalogs are public and cited in \S\ref{sec:method}.

\bibliographystyle{ACM-Reference-Format}
\bibliography{references}

\end{document}